\documentstyle[emulateapj,psfig]{article}
\def\insertplot#1#2#3#4#5#6#7{
\vskip 10pt\nobreak\hbox to \hsize{\hss\dimen0=#3in\hbox to #6\dimen0{%
\dimen0=#2in\vbox to #6\dimen0{\vss
\special{ps: plotfile #1}
\special{ps::[end]
  PGPLOT restore
}
}\hss}\hss}\vskip 10pt}

\slugcomment{To appear in the Astrophysical Journal Supplement, Vol. 154, September, 2004 issue}

\begin{document}
\title{The Formation and Evolution of Planetary Systems: \\
First Results from a Spitzer Legacy Science Program}
\author{M.R. Meyer\altaffilmark{1}, 
L.A. Hillenbrand\altaffilmark{2}, 
D.E. Backman\altaffilmark{3}, 
S.V.W. Beckwith\altaffilmark{4,13}, 
J. Bouwman\altaffilmark{5}, 
T.Y. Brooke\altaffilmark{2}, 
J.M. Carpenter\altaffilmark{2},
M. Cohen\altaffilmark{6},
U. Gorti\altaffilmark{3},
T. Henning\altaffilmark{5},
D.C. Hines\altaffilmark{7},
D. Hollenbach\altaffilmark{3},
J.S. Kim\altaffilmark{1},
J. Lunine\altaffilmark{8},
R. Malhotra\altaffilmark{8},
E.E. Mamajek\altaffilmark{1},
S. Metchev\altaffilmark{2},
A. Moro--Martin\altaffilmark{1},
P. Morris\altaffilmark{9},
J. Najita\altaffilmark{10},
D.L. Padgett\altaffilmark{9},
J. Rodmann\altaffilmark{5},
M.D. Silverstone\altaffilmark{1},
D.R. Soderblom\altaffilmark{4}, 
J.R. Stauffer\altaffilmark{9},
E.B. Stobie\altaffilmark{1},
S.E. Strom\altaffilmark{10},
D.M. Watson\altaffilmark{11}, 
S.J. Weidenschilling\altaffilmark{12},
S. Wolf\altaffilmark{5}, 
E. Young\altaffilmark{1}, 
C.W. Engelbracht\altaffilmark{1}, K.D. Gordon\altaffilmark{1}, K. Misselt\altaffilmark{1}, 
J. Morrison\altaffilmark{1}, J. Muzerolle\altaffilmark{1}, and K. Su\altaffilmark{1}.}
\altaffiltext{1}{Steward Observatory, The University of Arizona, Tucson}
\altaffiltext{2}{Astronomy, California Institute of Technology, Pasadena, CA}
\altaffiltext{3}{NASA--Ames Research Center, Moffet Field, CA} 
\altaffiltext{4}{Space Telescope Science Institute, Baltimore, MD} 
\altaffiltext{5}{Max--Planck--Institut f\"ur Astronomie, Heidelberg, Germany} 
\altaffiltext{6}{Radio Astronomy, University of California, Berkeley, CA} 
\altaffiltext{7}{Space Science Institute, Boulder, CO} 
\altaffiltext{8}{Lunar Planetary Lab, The University of Arizona, Tucson, AZ}
\altaffiltext{9}{Spitzer Science Center, Caltech, Pasadena, CA} 
\altaffiltext{10}{National Optical Astronomy Observatory, Tucson, AZ}
\altaffiltext{11}{Physics and Astronomy, Univ. of Rochester, Rochester, NY} 
\altaffiltext{12}{Planetary Science Institute, Tucson, AZ} 
\altaffiltext{13}{Johns Hopkins University, Balitmore, MD} 

\begin{abstract} 
We present 3-160 $\mu$m photometry obtained with the IRAC and MIPS 
instruments for the first five 
targets from the Spitzer Legacy Science Program ``Formation and 
Evolution of Planetary Systems" and 4-35 $\mu$m 
spectro-photometry obtained with the IRS for two sources. 
We discuss in detail our observations of the debris disks 
surrounding HD 105 (G0V, 30 $\pm 10$ Myr) and  
HD 150706 (G3V, $\sim$ 700 $\pm 300$ Myr).  
For HD 105, possible interpretations include large bodies 
clearing the dust inside of 45 AU or
a reservoir of gas capable of sculpting the dust distribution. 
The disk surrounding HD 150706 also exhibits 
evidence of a large inner hole in its dust distribution. 
Of the four survey targets without previously detected IR excess,
spanning ages 30 Myr to 3 Gyr, the new detection of excess 
in just one system of intermediate age suggests a variety of 
initial conditions or divergent evolutionary paths for debris 
disk systems orbiting solar-type stars.
\end{abstract}

\keywords{ stars, circumstellar disks, planet formation}

\section {Introduction}

A combination of optical, infrared, and millimeter observations 
has provided incontrovertible evidence over the past two decades
that most stars are surrounded at birth by 
circumstellar accretion disks (e.g. Beckwith and Sargent, 1996).  
That at least some of these disks build planets 
has become clear from radial velocity and photometric studies 
revealing M~sin$i$=0.2-15 M$_J$ 
planets orbiting nearby stars (e.g. Marcy et al. 2000). 
More indirect but still compelling evidence of planet 
formation comes from optical and infrared imaging of a few debris disks
(e.g. Kalas, Liu, \& Mathews 2004 and Weinberger et al. 1999). 
These solar system-sized dust disks 
are comprised of micron-sized grains produced as by-products of collisions 
between asteroid-like bodies with orbits expected to be dynamically 
stirred by massive planets (e.g. Lagrange et al. 2000). 

Following the IRAS discovery of excess infrared emission associated with
Vega (Auman et al. 1984), IRAS and ISO identified several dozen debris disks 
around fairly luminous main sequence stars.  Neither observatory, however,
had the sensitivity to detect {\it solar-type stars} 
down to photospheric levels at distances greater than a few pc, 
up to now preventing a complete census of solar-system-like debris disks over
a wide ranges of stellar ages. 
The extant samples have made it difficult
to infer the typical path of debris disk evolution 
(Habing et al. 1999; Meyer and Beckwith, 2000; Spangler et al. 2001).  
The unprecedented sensitivity of the Spitzer Space Telescope (Spitzer; 
Werner et al. 2004) will enable the Legacy Science Program 
{\it The Formation and Evolution
of Planetary Systems:  Placing Our Solar System in Context}
to search for debris systems around 330 stars with spectral types 
F8V to K3V and ages ranging from 3 Myr to 3 Gyr.  
Observations and results described here are from early validation
data taken with all three Spitzer instruments.
Targets were selected from among
those visible during the validation campaign of December 2003 
to be representative of the overall FEPS sample, the primary control 
variable of which is stellar age. Observed were HD 105,
HD 47875, HD 150706, HD 157664, and HD 161897 ranging in 
age from 30 Myr to 3 Gyr.  
We report here photometry for all of these stars, 
and low resolution spectro-photometry for two with 
observed IR excess, HD 105 and HD 150706. 

Stellar properties for our validation sample are summarized here 
and reported in Table 1. 
HD 105 (G0V, Houk 1978; 40 $\pm$ 1 pc, Hipparcos) is a kinematic
member of the Tuc-Hor moving group (Mamajek et al. 2004).  It is a
coronally and chromospherically active dwarf star with youth
established through its Li I $\lambda$6707 equivalent width
(e.g. Wichmann et al. 2003), its position
in the HR diagram, and activity indicators such as 
X-ray emission (Cutispoto et al. 2002) and 
Ca II HK emission (Wright et al. 2004). 
We adopt an age of 30 $\pm 10$ Myr for HD 105. 
HD 150706 (G3V, Buscombe 1998; 27 $\pm$ 0.4 pc, Hipparcos)
is an active main sequence dwarf 
with chromospheric Ca II HK emission 
(Wright et al. 2004) suggesting an age of 
635-1380 Myr adopting the calibration of Donahue et al. (1996)
and coronal X-ray emission (Voges et al 1999).  
The Li I $\lambda$6707 equivalent width (e.g. Soderblom \& Mayor 1993)
indicates an age consistent with 
possible kinematic membership in the UMa group 
(e.g. King et al.  2003) of $\sim$300-500 Myr.
We adopt the weighted average of these two age estimates, 
700 $\pm 300$ Myr for HD 150706.  HD 47875 is a young active 
X-ray emitting star (Favata et al. 1995) with 
kinematics consistent with membership 
in the local association.   We assign an age of 30--200 Myr
to this source.  HD 157664 is a galactic disk field star.
Because there are no indications of 
youth and because volume--limited samples of 
sun--like stars are most likely 1--3 Gyr old 
(e.g. Rocha-Pinto et al. 2000), we tentatively assign
this age to HD 157664.  HD 161897 is similarly inactive 
and we assign a preliminary age based on its Ca II
H \& K emission (Wright et al. 2004) of 1--3 Gyr.

\section{Spitzer Space Telescope Data} 

Next we describe the data acquisition and reduction
strategies for each instrument.   The derived flux
densities for all five sources are presented in Table 2. 

IRAC (Fazio et al. 2004) observations in each of the four channels
used the 32x32 pixel sub--array mode 
with an effective integration time of 0.01 sec per image 
(frame-time of 0.02 sec). The 64 images at each position in
the four--point--random dither pattern provided a total 
integration time of 2.56 sec per channel. 
We began with the Basic Calibrated Data (BCD) products of the Spitzer
Science Center (SSC) S9.1 data pipeline as described 
in the Spitzer Observer's Manual v4.0 
(hereafter SOM\footnote{http://ssc.spitzer.caltech.edu/documents/som/}). 
Aperture photometry was performed using IDP3
(Schneider \& Stobie 2002) v2.9. 
We used a 2-pixel radius aperture centered on the target and estimated
background beyond an 8-pixel radius as the median of $\sim$820 pixels. 
Background flux was normalized to the area of the target aperture 
and subtracted from the summed target flux.
The final source flux is the median of the 256 measures,
corrected from a 2-pixel radius to the 10-pixel radius
used for the IRAC instrumental absolute flux calibration.
Measurement uncertainty was estimated as the standard error in the mean and
added in quadrature to an absolute flux calibration uncertainty of 10\%.

Low resolution ($\lambda/\Delta\lambda \ \simeq$ 70 -- 120) spectra 
were obtained with the IRS (Houck et al. 2004) 
over the entire wavelength range available (5.2--38 $\mu$m) for all validation 
targets. 
We present here reduced spectral observations in Figure 1 for 
HD 105 and HD 150706 only.  
We used an IRS high--accuracy blue peak--up to acquire the source in 
the spectrograph slit.  Integration times per exposure were 
6 sec over the short-low wavelength range (5.2--14.5 $\mu$m), 
and either 6 sec (HD 150706) or 14 sec (HD 105) 
over the long-low wavelength range (14.0--38.0 $\mu$m). 
One cycle, resulting in spectra at two nod positions, was obtained in
staring mode for averaging and estimating the noise.
The BCDs resulting from the SSC pipeline S9.1 were reduced 
within the SMART software package (Higdon et al. 2004, in preparation).
We used the {\it droopres} data products before stray--light and 
flat--field corrections were applied.   
Spectra were extracted assuming point source profiles 
with a fiducial width of 5-6 pixels in the center of the orders but
allowing for variable width to account for increasing PSF size as a 
function of wavelength. Residual emission (mostly due to 
solar system zodiacal dust) 
was subtracted using adjacent pixels.  The background--subtracted
spectrum was divided by the spectrum of a 
photometric standard star ($\alpha$ Lac) 
and multiplied by an appropriately-binned template spectrum
for this standard provided by the IRS instrument team. 
Random errors calculated from 
the difference between the two independent spectra were
added in quadrature with an estimated 15 \% uncertainty in 
absolute flux calibration to produce the spectra shown in Figure 1. 

MIPS (Rieke et al. 2004) observations were obtained in all three bands 
using the small field photometry mode with
2 cycles of 3 sec Data Collection Events (DCEs) at 24 $\mu$m and 2 cycles
of 10 sec DCEs at 160 $\mu$m, approaching the confusion limit (Dole et al.
2004).  At 70$\mu$m, we observed HD 105, HD 150706 and HD 161897
for 2 cycles, HD 157664 for 4 cycles and HD 47875 for 7 cycles
all with 10 sec DCEs. 
After initial processing by the SSC S9.1 pipeline to provide reconstructed
pointing information, the MIPS data were further reduced using the MIPS Data
Analysis Tool (DAT, v2.71) developed by the MIPS Instrument Team (Gordon et
al. 2004).  This includes the ``enhancer" portion of the DAT,
which corrects for distortion in individual images and 
combines them onto a sub-sampled tangential plane mosaic. 
We present images of HD 105 and 
HD 150706 at 70 and 160 $\mu$m in Figure 2. 
Aperture photometry was performed in IDP3 
with target apertures of 14.99", 29.70", and 47.8" at 
24, 70, and 160 $\mu$m,
respectively.  We used background annuli of 29.97--42.46'' for 
24 $\mu$m, 39.6--79.2'' for 70 $\mu$m, and 47.9--79.8'' for 160 $\mu$m. 
The mean background per pixel was scaled to the appropriate aperture size and
subtracted from the summed flux inside each aperture.
Random uncertainties were determined from the ensemble of measurements
for the 24 $\mu$m observations, and from the noise in the background
for the mosaicked images at 70 and 160 $\mu$m.  These error estimates
were added in quadrature with uncertainties in the absolute calibration
of 10, 20, and 40 \% for 24, 70, and 160 $\mu$m respectively as measured
for sources as faint at 60 mJy as 70 $\mu$m. 
Upper limits were derived for sources not detected based on photometry attempted
at the source position estimated from the coordinates given in the image
headers.  Three times the estimated noise was 
used to determine an upper limit if the 
inferred SNR of the measured flux was $<$ 3$\sigma$. 

\section{Spectral Energy Distributions of Targets} 

We present SEDs for HD 105 and HD 150706 in Figure 1. 
We have adopted the recommended (SOM) 
central wavelengths and have not applied color--corrections
which are still uncertain and much smaller than 
the quoted absolute calibration uncertainties. 
The photospheric emission component was modeled by fitting
Kurucz atmospheres including convective overshoot to available
$BV$ Johnson, $vby$ Stromgren, $B_TV_T$ 
Tycho, $H_p$ Hipparcos, $RI$ Cousins, and $J$, $H$, $K_s$ 2MASS photometry. 
Predicted magnitudes were computed as described in 
Cohen et al. (2003,and references therein)
using the combined system response of filter,
atmosphere (for ground-based observations), and detector. 
The best-fit Kurucz model was computed in a least squares sense with the 
effective temperature and normalization constant (i.e. radius) as free 
parameters, [Fe/H] fixed to solar metallicity, and surface gravity fixed to 
the value appropriate for the adopted stellar age and mass. Visual 
extinction was fixed to $A_V = 0^m$ for stars with distances 
less than 40~pc, assumed to be within the dust--free Local Bubble, but 
a free parameter for HD~47875 and HD~157664.
The adopted stellar parameters are listed in Table 1. 

Emission in excess of that expected from the stellar photospheres is
found in two of our five validation targets,  HD 105 and HD 150706. 
The remaining three validation targets have no significant
excess emission and were not detected at 70 $\mu$m
despite sensitivities comparable to the  HD 105 and HD 150706
observations.  We place limits on the ratio $L_{IR}/L_{*}$ 
for these three stars based on the observed upper limits
(assuming a dust temperature of $\sim$ 40 K) in Table 1 
and discuss the two excess sources in detail. 

HD 105 was found to have an IR excess by Silverstone
(2000) based on 60 and 90 $\mu$m ISO/ISOPHOT measurements.
We confirm that with Spitzer measurements of excess at 70 and
160 $\mu$m but find no obvious excess at $\lambda$ $<$ 35 $\mu$m 
based on the IRS spectra or IRAC/MIPS photometry. 
The total flux in the excess calculated via trapezoidal
integration from 24 to 1200
microns is $\sim 1 \times 10^{-14}$ W m$^{-2}$, 
corresponding to $L_{IR}/L_{*}$ = $\sim 3.9 \times 10^{-4}$.
Assuming a temperature of $\sim$ 40 K, the estimated solid
angle subtended by total effective particle cross-section is $\sim 2
\times 10^{-13}$ sr = 13 AU$^2$ for d =
40 pc.  Based on the 24 $\mu$m measurement, we estimate
there can be no more than 
$3 \times 10^{-3}$ as much radiating area at 100 K, and 
$<$ $4 \times 10^{-5}$ as much at 300 K, as there is at 40 K. 
HD 150706 has a newly discovered IR excess at 70 $\mu$m, but 
only an upper limit at 160 $\mu$m and no
evidence for excess at $\lambda$ $<$ 35 $\mu$m.  The 
total flux in the excess is $\sim 3 \times 10^{-15}$ W
m$^{-2}$, corresponding to fractional IR luminosity $\sim 5.4
\times 10^{-5}$.   Assuming a temperature $<$ 84 K, the total
effective particle cross-section is $> 0.09$ AU$^{2}$ for d
= 27 pc. 

\section{Disk Properties and Interpretation} 

Following arguments 
employed for Vega and the other main sequence debris disk archetypes
discovered by IRAS (Backman and Paresce 1993), 
we assume the IR excess emission around HD 105 and
HD 150706 is from grains orbiting, and in thermal
equilibrium with radiation from, the central stars.  
Model inner and outer radii for disks containing the cold material around
HD 105 and HD 150706 can be calculated with assumptions regarding
grain composition, size distributions, and spatial distributions.
The lack of distinct mineralogical features in the observed IRS
spectra (which would constrain the dust properties) 
means there can be no unique model but rather a
range of models that satisfy the observations.
For HD 150706 the single data point for the IR excess
translates into a limit that the material must lie farther from the
star than $\sim$ 11 AU if it is in the form of ``blackbody" grains
larger than the longest wavelength of significant emission.  Smaller
grains would satisfy the same temperature constraint at larger
distances from the star.  
The material around HD 105 is consistent with being distributed 
in a narrow ring with inner edge R$_{IN}$
at 42 $\pm$ 6 AU and an outer edge at R$_{OUT}$ -  R$_{IN}$ $<$ 4 AU if
``blackbody" grains are assumed.  The ranges result from
photometric uncertainty and are independent of the assumed grain surface 
density radial power law exponent within the range
$\Sigma (r) \sim r^{[-2.0,0]}$.  
If material in the inner ``hole" is assumed to have constant
surface density with radius (as would be produced through the 
Poynting--Robertson (P--R) effect), the surface 
density in the zone at r $<$ R$_{IN}$ is less than $3 \times 10^{-2}$ of the
model surface density at r $>$ R$_{IN}$.
Another family of models containing intermediate-sized 
(graybody) grains with
emissivity falling as $1/\lambda$ beyond $\lambda = 40 \mu$m possesses 
inner edges R$_{IN}$ ranging from about 50 to 70 AU and outer edges
R$_{OUT}$ ranging from 250 to 1500 AU depending on the assumed radial power
law exponent of the surface density distribution. 

Given the above results for HD 105 and HD 150706 
from simple models with strong 
(but reasonable) assumptions about the grain properties
of the observed disks, we now explore ranges of disk 
models that are consistent with the data following 
Wolf \& Hillenbrand (2003).  For grain compositions we assumed 
``astronomical" silicate plus graphite in the ISM ratio 
and 
surface density distribution $\Sigma(r) \propto r^0$. 
The mass of the disk was adjusted to match 
the peak flux in the infrared excess.
Parameters such as grain size distribution $n(a) \sim a^{-p}$
power--law exponent, minimum/maximum grain size, 
and the inner/outer edge of the disk, were varied to find
the range of values 
consistent with the observed spectral energy distribution
of HD 105. 
The models were relatively insensitive 
to the radial density distribution 
exponent. 
The wavelength at which the dust re-emission spectrum begins to
depart significantly from the stellar photosphere was used to find
the smallest grain size and smallest inner disk radius consistent with
the data.  These two parameters are degenerate resulting in 
single grain sizes in the range 0.3, 5, and 8 $\mu$m 
requiring inner gap sizes of
1000, 120, and 42 AU respectively.  Adopting a minimum grain size of 5 $\mu$m
 and allowing 
for a grain size distribution up to 100 or 1000 $\mu$m produced 
lower $\chi^2$ fits and decreased the required inner radius from 120 to 45 AU
(32 AU for a$_{MIN}$ $\sim$ 8 $\mu$m or larger). 
The upper grain size, if one 
exists, and the outer radius are not well constrained in the absence 
of sub-millimeter measurements. The mass in grains $<$1 mm for the above
models is between 9$\times10^{-8}$ and 4$\times10^{-7}$ M$_\odot$.
For HD 150706, we used the measured 70 $\mu$m flux
and 160 $\mu$m upper limit and the methodology outlined 
above to model the disk.   Compared to HD 105, we find
a smaller minimum grain size (0.3 or 1 $\mu$m) and 
a narrower disk (inner radius $\sim$ 45 or $\sim$ 20 AU 
respectively with outer radius $<$ 100 AU). 
For the remaining three objects, 
we used the best--fit model for HD 150706, and
scaled the results to the observed upper--limits on
$L_{IR}/L_*$ to 
derive the dust mass upper limits given in Table 1. 

In the cases of our two disk detections,
assuming a grain density
\footnote{This density is appropriate for 
solid silicate dust grains which we 
assume are the result of comminuted 
Kuiper Belt Object or cometary analogs 
whose bulk densities are lower.}  of 
2.5 g/cm$^3$
we can calculate the P--R drag timescale.
For HD 105 removal of 5 $\mu$m grains occurs $<$ 15 Myr at 45 AU
compared to a stellar age of 30 Myr, suggesting that any such small grains
are regenerated, perhaps through collisions of planetesimals.
However, given the optical depth of the dust
(15--300 AU$^2$, the radiating cross--sectional area), the
timescale for dust grains to collide is $<$ 10$^6$ yrs
suggesting that collisions as well as P--R drag are important
in determining the actual size distribution of the dust as well
as its radial surface density profile.  
For HD 150706 the P--R drag timescale for 1 $\mu$m grains at 20 AU is
$<$ 1 Myr compared to an age of 700 $\pm$ 300 Myr suggesting 
regeneration of dust through collisions of planetesimals as
in the case of HD 105. 
In both cases the lack of circumstellar material in the 
inner disk (expected from a model of P--R drag) suggests:
1) something is preventing dust at 20--45 30 AU from reaching the sublimation
radius in the inner disk; and 2) a lack of 
significant numbers of colliding planetesimals inside of 20--45 AU. 
This suggests that the inner region is relatively clear of small bodies, 
consistent with some estimates for the timescale of terrestrial planet 
formation (e.g. Kenyon and Bromley, 2004). 
The presence of one or more large planets interior to $\sim$ 20--45 
AU may explain the inner edge of the outer dust disk 
(e.g. Moro--Martin and Malhotra, 2003). 
The preceeding discussion presumed that the gas to dust mass ratio 
is $<$ 0.1, for which the dust dynamics are driven 
by interactions with the radiation
field of the central star.  Our high resolution
IRS observations of the $\sim$ 30 Myr star HD 105, 
still under analysis, should be sensitive to small 
amounts of gas between 50--200 K.  
The presence of a remnant inner gas disk  
can influence dust migration and produce a ring morphology 
such as we infer for the dust (Takeuchi and Artymowicz, 2001). 

HD 150706, with an age of 700 $\pm$ 300 Myr, is less
likely to retain a gas--rich disk than HD 105 
(e.g. Zuckerman et al. 1995).  However, it does exhibit 
evidence for a previously undetected dust disk 
while HD 47875, HD 157664, and HD 161897, ranging in age 
from 30 Myr to 3 Gyr, do not at levels comparable to 
the HD 150706 detection (Table 1).   The 
disk surrounding HD 150706 has a hole devoid of dust 
with inner radius of at least 20 AU, 
comparable to the disk surrounding HD 105. 
Perhaps the inner region of this disk is 
being kept clear by the presence of 
larger bodies as discussed above. 
While we cannot draw robust conclusions from these small samples, 
it is clear that other factors (range in primordial disk 
properties and/or variations in evolutionary histories?) 
are required to explain the growing body of
debris disk observations besides a simple monotonic decrease 
in dust mass with age. 

\section{Acknowledgements}

We would like to thank our colleagues at the Spitzer
Science Center and members of the instrument teams, 
for their help in analyzing the Spitzer data. 
FEPS is pleased to acknowledge support
through NASA contracts 1224768, 1224634, and 1224566 administered through JPL.

\vfill\eject
\begin{figure}
\insertplot{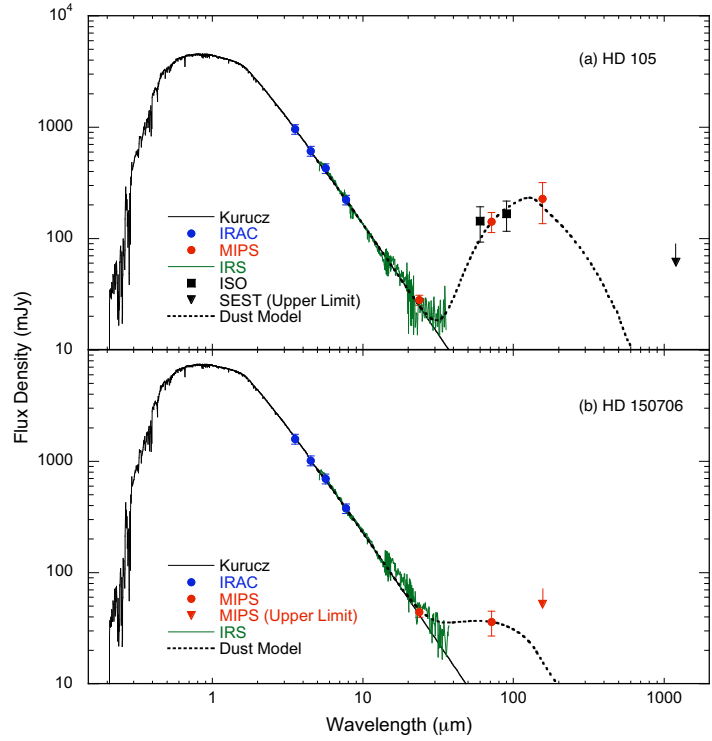}{4.0}{5.0}{2.5}{0.0}{0.50}{0}
\caption{
Spectral energy distributions for HD 105 and HD 150706 
using data from Table 2 and stellar models from Table 1.  
Also shown are representative 
models for the dust debris disks surrounding HD 105 
($R_{IN}/R_{OUT}=45/300 AU$; $a_{min}/a_{max} = 5/100 \mu m$) 
and HD 150706 
($R_{IN}/R_{OUT}=20/100 AU$; $a_{min}/a_{max} = 1/100 \mu m$) 
discussed in the text. 
} 
\label{sed}
\end{figure}

\begin{figure}
%\insertplot{f2.ps}{4.0}{5.0}{2.5}{0.0}{0.50}{0}
\caption{MIPS 70 and 160 $\mu$m images from which photometry was
derived.  These images have been four times over-sampled. 
[NOTE:  See separate file for download.]} 
\label{sed}
\end{figure}

\vfill\eject

\begin{center}
\begin{table}
\begin{small}
\caption{Adopted Stellar and Derived Circumstellar Properties}
\resizebox{500pt}{!}{
\begin{tabular}{lrrrrccrrrrr}
\multicolumn{1}{c}{Name}   &
\multicolumn{1}{c}{SpT} &
\multicolumn{1}{c}{$T_{eff}$/K}$^1$ &
\multicolumn{1}{c}{$log(L_{*}/L_{\odot}$)}$^1$ &
\multicolumn{1}{c}{log(g)}$^1$ &
\multicolumn{1}{c}{$A_{V}$/mag}$^1$ &
\multicolumn{1}{c}{$\tau$/Myr}$^2$ &
\multicolumn{1}{c}{d/pc}$^2$ &
\multicolumn{1}{c}{log $L_{IR}/L_*$}&
\multicolumn{1}{c}{M$_{DUST}$/M$_{\odot}$}$^3$&
\multicolumn{1}{c}{R$_{IN}/AU$}$^3$&
\multicolumn{1}{c}{R$_{OUT}/AU$}$^3$ \\\hline\hline

HD 105    &  G0V  &  6063 & 0.11 & 4.47 &0.00    &  30 $\pm$ 10  & 40$\pm$1  & -3.41 &  $\sim 1.0 \times 10^{-7}$ &  45.0 & --- \\
HD 47875  &  G3V  &  5886 & 0.10 & 4.61 &0.36   &  30--200  & 70  & $<$ -3.85 &  $<  1.8 \times 10^{-7}$ &  --- & ---  \\ 
HD 150706 &  G3V &  5958  & -0.03 & 4.62 &0.00 & 700 $\pm$ 300 & 27$\pm$0.4  & -4.27 &  $\sim 6.9 \times 10^{-8}$ &  20 & $<$100 \\
HD 157664 & G0  & 6494 & 0.63 & 4.50 &0.18& 1000--3000   & 84$\pm$5  &  $<$ -4.22 &  $< 6.1 \times 10^{-8}$ &  --- & --- \\
HD 161897 & K0  & 5579 & -0.18 & 4.50 &0.00& 1000--3000  & 29$\pm$1  & $<$ -4.31 &  $< 6.3 \times 10^{-8}$ &  --- &  ---\\\hline 
\tablecomments{\footnotesize 
\\
$^1$ Based on the photospheric model fitting proceedure described in the text. 
\\
$^2$ Derived from data taken from the literature.
\\
$^3$ Based on the circumstellar disk models described in the text.
}
\label{props}
\end{tabular}
}
\end{small}
\end{table}
\end{center}

\begin{center}
\begin{table}
\begin{small}
\caption{Spitzer Photometry in milli--Jy} 

\resizebox{500pt}{!}{
\begin{tabular}{lrrrrrrrrrrrrrr}
\multicolumn{1}{c}{Source } &
\multicolumn{1}{c}{3.6 $\mu$m} & 
\multicolumn{1}{c}{$\sigma$} & 
\multicolumn{1}{c}{4.5 $\mu$m} &
\multicolumn{1}{c}{$\sigma$} &
\multicolumn{1}{c}{5.8 $\mu$m} &
\multicolumn{1}{c}{$\sigma$} &
\multicolumn{1}{c}{8.0   $\mu$m} &
\multicolumn{1}{c}{$\sigma$} &
\multicolumn{1}{c}{24 $\mu$m} &
\multicolumn{1}{c}{$\sigma$} &
\multicolumn{1}{c}{70 $\mu$m} & 
\multicolumn{1}{c}{$\sigma$} &
\multicolumn{1}{c}{160 $\mu$m} & 
\multicolumn{1}{c}{$\sigma$} \\\hline\hline

HD 105    & 960   &  96  &  610   & 61  &  427   &  43 &  222 &  22 & 28.0 & 3.0 & 142 & 29 & 227  & 91\\
HD 47875  &  327  & 33  &  210    & 21  &   153  &  16 &  77  &   8 & 9.4 & 1.1 & $<$ 18 & NA & $<$ 42 & NA \\ 
HD 150706 & 1587  & 159  &  1014  & 102 &  697   &  70 &  377 &  38 & 44.2 & 4.5 & 36 & 9 & $<$ 52 & NA \\
HD 157664 & 582   & 58   &  368   & 37  &  252   &  25 &  133 &  13 & 15.0 & 1.7 & $<$ 18 & NA & $<$ 36 & NA \\
HD 161897 & 1142  & 114  &  715   & 72  &  495   &  50 &  266 &  27 & 29.5 & 3.1  & $<$ 19 & NA & $<$ 35 & NA  \\ \hline
\tablecomments{\footnotesize 
Fluxes quoted are based on the calibrations used
in the SOM v4.0.
Errors quoted are derived as discussed in the
text including calibration uncertainties.  Upper
limits are 3 $\sigma$ as discussed in the text.} 
\label{phot}
\end{tabular}
}
\end{small}
\end{table}
\end{center}


\begin{references}
\reference{} 
Aumann, H.~H., et al.\ 1984, \apjl, 278, L23 
\reference{} 
Backman, D.~E.~\& 
Paresce, F.\ 1993, Protostars and Planets III, 1253 
\reference{} 
Beckwith, S.~V.~W.~\& Sargent, A.~I.\ 1996, \nat, 383, 139 
\reference{} 
Buscombe, W.\ 1998, VizieR Online Data Catalog, 3206, 0. 
\reference{} 
Cohen, M., Megeath, S. T., Hammersley, P. L., Mart\'in-Luis, F.,
\& Stauffer, J. 2003 \aj, 125, 2645
\reference{} 
Cutispoto, G., 
Pastori, L., Pasquini, L., de Medeiros, J.~R., Tagliaferri, G., \& 
Andersen, J.\ 2002, \aap, 384, 491 
\reference{} 
Dole, H. et al. 2004, ApJ, submitted. 
\reference{} 
Donahue, R.~A., Saar, S.~H., \& Baliunas, S.~L.\ 1996, \apj, 466, 384 
\reference{} 
Favata, F., Barbera, M., Micela, G., \& Sciortino, S.\ 1995, \aap, 295, 147 
\reference{} 
Fazio, G. et al. 2004, ApJS, this issue. 
\reference{} 
Gordon, K.D.\ et al.\ 2004, \pasp, submitted
\reference{} 
Gorti, U. \& Hollenbach, D. 2004, submitted. 
\reference{} 
Grogan, K., Dermott, S.~F., \& Durda, D.~D.\ 2001, Icarus, 152, 251 
\reference{} 
Habing, H.~J., et al.\ 1999, \nat, 401, 456 
\reference{} 
Houck, J., et al. 2004, ApJS, this issue. 
\reference{} 
Houk, N.\ 1978, Ann Arbor : 
Dept.~of Astronomy, University of Michigan : distributed by University 
Microfilms International, 1978. 
\reference{} 
Kalas, P., 
Liu, M.~C., \& Matthews, B.~C.\ 2004, Science, 303, 1990 
\reference{} 
Kenyon, S.~J.~\& Bromley, B.~C.\ 2004, \apj, 602, L133
\reference{} 
Lagrange, A.-M., Backman, D.~E., \& Artymowicz, P.\ 2000, 
Protostars and Planets IV (Tucson: University of Arizona Press; 
eds Mannings, V., Boss, A.P., Russell, S. S.), 639
\reference{} 
King, J.~R., Villarreal, 
A.~R., Soderblom, D.~R., Gulliver, A.~F., \& Adelman, S.~J.\ 2003, \aj, 
125, 1980 
\reference{} 
Mamajek, E.E., et al. 2004, ApJ, submitted. 
\reference{} 
Marcy, G.~W., Cochran, W.~D., \& Mayor, M.\ 2000, Protostars and Planets IV, 1285 
\reference{} 
Meyer, M.~R.~\& 
Beckwith, S.~V.~W.\ 2000, ISO Survey of a Dusty Universe, 
Edited by D.~Lemke et al., Lecture Notes 
in Physics, vol.~548, p.341, 341 
\reference{} 
Moro-Mart{\'{\i}}n, A.~\& Malhotra, R.\ 2003, \aj, 125, 2255 
\reference{} 
Rieke, G.H. et al. 2004, ApJS, this issue. 
\reference{} 
Rocha-Pinto, H.~J., Maciel, W.~J., Scalo, J., \& Flynn, C.\ 2000, \aap, 
358, 850 
\reference{} 
Schneider, G., \& Stobie, E. 2002, ASP Conf. Ser. 281, ADASS 
XI, ed. D. A. Bohlender, D. Durand, and
T. H. Handley (San Francisco: ASP), p. 382
\reference{}
 Silverstone, M.~D.\ 2000, Ph.D.~Thesis,  UCLA. 
\reference{}
Soderblom, D.~R., Pilachowski, C.~A., 
Fedele, S.~B., \& Jones, B.~F.\ 1993, \aj, 105, 2299 
\reference{}
Spangler, C., Sargent, 
A.~I., Silverstone, M.~D., Becklin, E.~E., \& Zuckerman, B.\ 2001, \apj, 
555, 932  
\reference{} 
Takeuchi, T.~\& 
Artymowicz, P.\ 2001, \apj, 557, 990 
\reference{} 
 Voges, W., et al.\ 1999, 
VizieR Online Data Catalog, 9010, 0 
\reference{} 
Weinberger, A. et al. \ 1999, ApJ, 525, L53 
\reference{} 
Werner, M.W. et al. 2004, ApJ, this issue. 
\reference{} 
Wichmann, R., Schmitt, J.~H.~M.~M., \& Hubrig, S.\ 2003, \aap, 399, 983 
\reference{} 
Wolf, S.~\& Hillenbrand, L.~A.\ 2003, \apj, 596, 603 
\reference{} 
Wright, J.~T., Marcy, G.~M., Butler, R.~P., \& Vogt, S.~S. 2004, ApJS 
in press (astro-ph/0402582). 
\reference{} 
Zuckerman, B., Forveille, T., \& Kastner, J.~H.\ 1995, \nat, 373, 494 
\end{references}
\end{document}